\documentclass[aps,prb,reprint,superscriptaddress,preprintnumbers]{revtex4-2}
\usepackage[T1]{fontenc}
\usepackage{times}
\usepackage{graphicx,color}
\usepackage{amsfonts,amsmath,amssymb,amsbsy}
\usepackage[colorlinks=true, linkcolor=blue, citecolor=blue, urlcolor=blue]{hyperref}

\let\mathbf=\boldsymbol

\def\emph#1{\textcolor{magenta}{#1}}
\begin{document}

\title{Current-induced dynamics of skyrmion tubes in synthetic antiferromagnetic multilayers}

\author{Jing Xia}
\thanks{These authors contributed equally to this work.}
\affiliation{College of Physics and Electronic Engineering, Sichuan Normal University, Chengdu 610068, China}
\affiliation{School of Science and Engineering, The Chinese University of Hong Kong, Shenzhen, Guangdong 518172, China}

\author{Xichao Zhang}
\thanks{These authors contributed equally to this work.}
\affiliation{Department of Electrical and Computer Engineering, Shinshu University, 4-17-1 Wakasato, Nagano 380-8553, Japan}

\author{Kai-Yu Mak}
\affiliation{School of Science and Engineering, The Chinese University of Hong Kong, Shenzhen, Guangdong 518172, China}

\author{Motohiko Ezawa}
\affiliation{Department of Applied Physics, The University of Tokyo, 7-3-1 Hongo, Tokyo 113-8656, Japan}

\author{Oleg A. Tretiakov}
\affiliation{School of Physics, The University of New South Wales, Sydney 2052, Australia}

\author{\\ Yan Zhou}
\email{zhouyan@cuhk.edu.cn}
\affiliation{School of Science and Engineering, The Chinese University of Hong Kong, Shenzhen, Guangdong 518172, China}

\author{Guoping Zhao}
\email{zhaogp@uestc.edu.cn}
\affiliation{College of Physics and Electronic Engineering, Sichuan Normal University, Chengdu 610068, China}

\author{Xiaoxi Liu}
\email{liu@cs.shinshu-u.ac.jp}
\affiliation{Department of Electrical and Computer Engineering, Shinshu University, 4-17-1 Wakasato, Nagano 380-8553, Japan}

\begin{abstract}
Topological spin textures can be found in both two-dimensional and three-dimensional nanostructures, which are of great importance to advanced spintronic applications. Here we report the current-induced skyrmion tube dynamics in three-dimensional synthetic antiferromagnetic (SyAF) bilayer and multilayer nanostructures. It is found that the SyAF skyrmion tube made of thinner sublayer skyrmions is more stable during its motion, which ensures that a higher speed of the skyrmion tube can be reached effectively at larger driving current. In the SyAF multilayer with a given total thickness, the current-induced deformation of the SyAF skyrmion tube decreases with an increasing number of interfaces; namely, the rigidity of the SyAF skyrmion tube with a given thickness increases with the number of ferromagnetic (FM) layers. For the SyAF multilayer with an even number of FM layers, the skyrmion Hall effect can be eliminated when the thicknesses of all FM layers are identical. Larger damping parameter leads to smaller deformation and slower speed of the SyAF skyrmion tube. Larger fieldlike torque leads to larger deformation and a higher speed of the SyAF skyrmion tube. Our results are useful for understanding the dynamic behaviors of three-dimensional topological spin textures and may provide guidelines for building SyAF spintronic devices.
\end{abstract}

\date{May 7, 2021}

\preprint{\textsl{\href{https://doi.org/10.1103/PhysRevB.103.174408}{Physical Review B 103, 174408 (2021)}}}

\maketitle

\section{Introduction}
\label{se:Introduction}

Nanoscale spin textures in magnetic materials may exhibit unique static and dynamic properties due to their topological structures~\cite{Bogdanov_1989,Roszler_NATURE2006,Nagaosa_NNANO2013,Muhlbauer_SCIENCE2009,Yu_Nature2010,Wiesendanger_Review2016,Finocchio_JPD2016,Kang_PIEEE2016,Kanazawa_AM2017,Wanjun_PHYSREP2017,Fert_NATREVMAT2017,Duine_NP2018,Everschor_JAP2018,Zhou_NSR2018,Zhang_JPCM2020,Gobel_Phys.Rep2020,Mochizuki_Review}. An exemplary topological spin texture is the skyrmion texture, which was theoretically predicted in 1989~\cite{Bogdanov_1989} and experimentally observed in 2009~\cite{Muhlbauer_SCIENCE2009}. The magnetic skyrmion has been extensively studied in the past decade due to its intriguing physical properties and broad potential applications in functional spintronic devices~\cite{Wiesendanger_Review2016,Finocchio_JPD2016,Kang_PIEEE2016,Kanazawa_AM2017,Wanjun_PHYSREP2017,Fert_NATREVMAT2017,Everschor_JAP2018,Zhou_NSR2018,Zhang_JPCM2020}. In particular, the magnetic skyrmion can be used as a nonvolatile information carrier in magnetic memory~\cite{Parkin_SCIENCE2008,Fert_NNANO2013,Iwasaki_NC2013,Sampaio_NNANO2013,Tomasello_SREP2014,Yan_NCOMMS2014} and logic computing~\cite{Xichao_SREP2015B,ZhangSL_SCIREP2015,Xing_PRB2016,Luo_NanoLett2018} applications that meet future commercial requirements, such as ultrahigh storage density and ultralow energy consumption.

Towards the applications of skyrmions in magnetic and spintronic devices, several different skyrmion-hosting systems, ranging from quasi-two-dimensional to three-dimensional structures, have been developed and investigated using a variety of theoretical and experimental methods~\cite{Wiesendanger_Review2016,Finocchio_JPD2016,Kang_PIEEE2016,Kanazawa_AM2017,Wanjun_PHYSREP2017,Fert_NATREVMAT2017,Everschor_JAP2018,Zhou_NSR2018,Zhang_JPCM2020,Reichhardt_NC2020,Rybakov_PRL2015,Lin_PRB2016,Zheng_NatNano2018,Sohn_PRB2019,Koshibae_SciRep2020,Kanazawa_PRB2017,Zhang_PNAS2018,Birch_NC2020,Mandru_NC2020,Seki_NC2020,Yu_Nature2010,Muhlbauer_SCIENCE2009,Mathur_AM2021}. For example, the existence of magnetic skyrmions was first realized in magnetic ultrathin films and bulk materials, where skyrmions are stabilized by Dzyaloshinskii-Moriya (DM) interactions~\cite{Muhlbauer_SCIENCE2009,Yu_Nature2010}. Recently, the community has further focused on skyrmions in ferromagnetic (FM) multilayers with interface-induced DM interactions, where both the magnitude of DM interaction and the thermal stability of skyrmions can be enhanced due to the multilayer nanostructure~\cite{Moreau-Luchaire_NNANO2016,Woo_NMATER2016,Yu_NANOLETT2016,Soumyanarayanan2017,ZhangSenfu_APL2018,Duong_APL2019,Mandru_NC2020,Yang_PRL2015,Boulle_NNANO2016,Dupe_NCOMMS2016}.

However, FM skyrmions, either in single- or multilayer films, may show the skyrmion Hall effect when they are driven by spin currents~\cite{Zang_PRL2011,Wanjun_NPHYS2017,Litzius_NPHYS2017}, which is a dynamic phenomenon associated with the topological nature of skyrmions and usually leads to the accumulation or destruction of skyrmions at sample edges~\cite{Xichao_NCOMMS2016,Xichao_PRB2016B,Wanjun_NPHYS2017,Litzius_NPHYS2017}. Hence, many strategies have been proposed to eliminate the skyrmion Hall effect for spintronic applications based on in-line motion of skyrmions~\cite{Xichao_NCOMMS2016,Xichao_PRB2016B,Tomasello_JPD2017,Tretiakov_PRL2016,Xichao_SREP2016,Woo_NC2018,Duine_NP2018,Hirata_NatNano2019,Dohi_NC2019,Legrand_NatMater2020,Xichao_PRB2016C}.
A very important strategy is to create and manipulate skyrmions in synthetic antiferromagnetic (SyAF) bilayer and multilayer nanostructures~\cite{Xichao_NCOMMS2016,Xichao_PRB2016B,Duine_NP2018,Dohi_NC2019,Legrand_NatMater2020,Tomasello_JPD2017,Hrabec_APL2020,Siddiqui_JAP2020}.

In fact, the topic of SyAF multilayers has been studied for many years, and a lot of progress has been achieved in describing the behaviors of SyAF domains~\cite{Hellwig_JMMM2007,Bran_PRB2009} and SyAF domain walls~\cite{Yang_NatNano2015,Yang_NatPhys2019}. The focus is shifting from domains and domain walls to skyrmions in recent years.
The SyAF skyrmions carry a net topological charge of zero and thus are free from the skyrmion Hall effect. For example, a bilayer SyAF skyrmion consists of two skyrmions with opposite topological charges, where the topological Magnus forces acting on the two skyrmions are identical in magnitude but opposite in direction~\cite{Xichao_NCOMMS2016,Xichao_PRB2016B}. Therefore, the Magnus forces are adequately canceled out and the bilayer SyAF skyrmion can move straightly along the driving force direction. Recent state-of-the-art experiments have demonstrated the stabilization~\cite{Legrand_NatMater2020} and current-driven motion~\cite{Dohi_NC2019} of bilayer SyAF skyrmions at room temperature.

In thick SyAF multilayer structures, the SyAF skyrmion is more like a three-dimensional tube instead of a two-dimensional object.
Namely, the SyAF skyrmion tube can be seen as a stack of two-dimensional skyrmions aligned along the $z$ axis. It has some similarity to the pancake vortices in layered superconductors, where the system can be viewed as a collection of two-dimensional vortices in each plane coupled together~\cite{Clem_PRB1991}.
Note that similar pancake vortex effects were also observed experimentally in synthetic antiferromagnets~\cite{Oleg_SR2018}.
If the multilayer SyAF skyrmion consists of an even number of antiferromagnetically exchange-coupled skyrmions, the total skyrmion number of the SyAF skyrmion tube is equal to zero, and the skyrmion Hall effect can be eliminated in principle~\cite{Xichao_PRB2016B,Tomasello_JPD2017}. However, a large driving force may result in the distortion of the skyrmion tube in the thickness dimension and may further lead to more complex dynamic behaviors of the skyrmion tube~\cite{Kagawa_NC2017,Yokouchi_SciAdv2018}. Although the dynamics of the FM skyrmion tube have been studied in recent years~\cite{Leonov_PRL2016,Yokouchi_SciAdv2018,Sohn_PRB2019,Birch_NC2020,Koshibae_SciRep2020,Seki_NC2020,Zhang_PRL2018,Kagawa_NC2017,Mandru_NC2020,Mathur_AM2021}, the complex dynamics of a SyAF skyrmion tube still remains elusive. In this work, we systematically study the current-induced dynamics of skyrmion tubes in SyAF multilayers using both theoretical and computational approaches.

\begin{figure}[t]
\centerline{\includegraphics[width=0.490\textwidth]{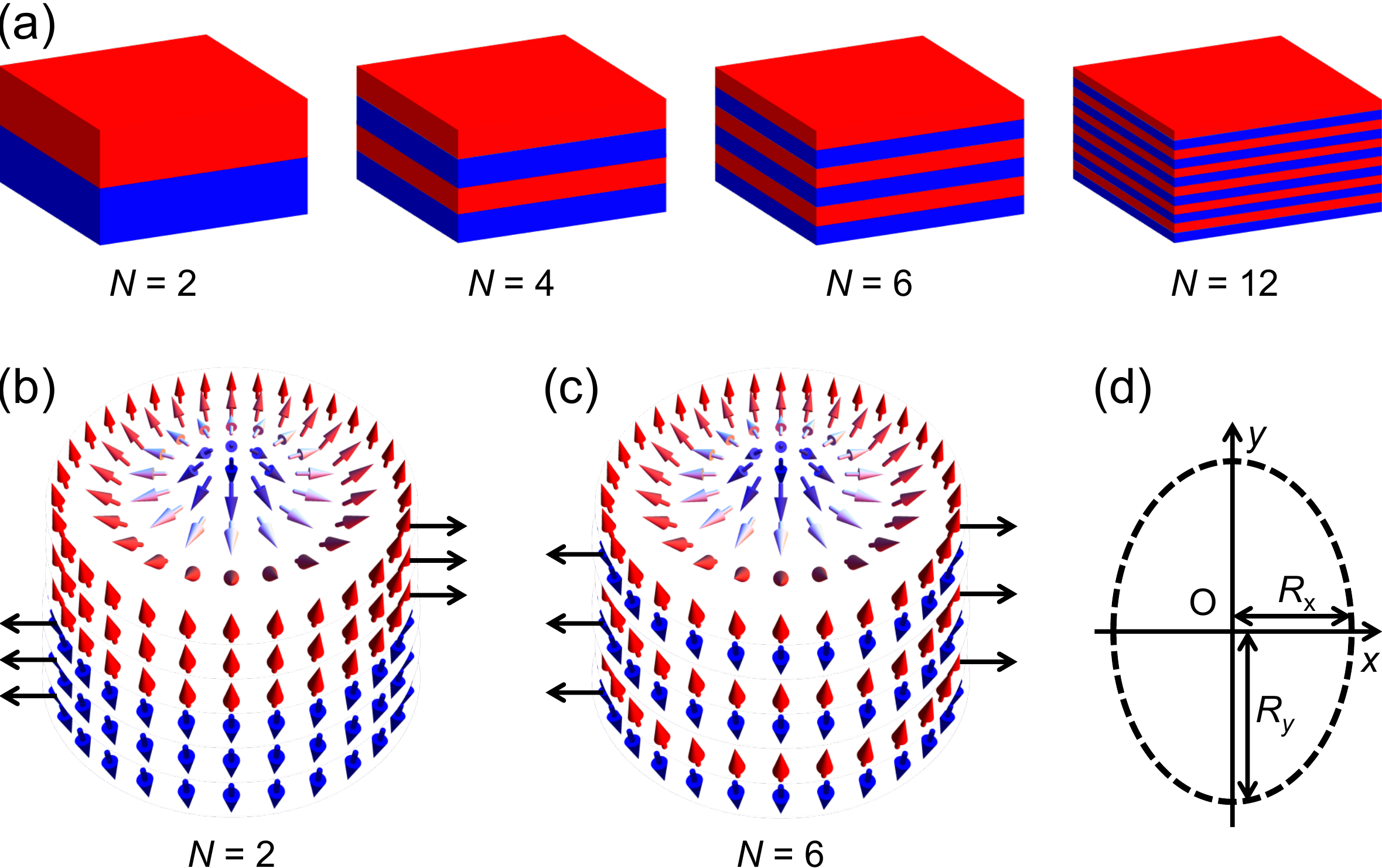}}
\caption{%
(a) Schematics of the simulation models. The total sample thickness is fixed at $12$ nm. $N$ denotes the number of FM layers in a sample. For $N=2$, the thickness of each FM layer equals $6$ nm. For $N=12$, the thickness of each FM layer equals $1$ nm. In each sample, the adjacent FM layers are antiferromagnetically exchange coupled, forming a SyAF structure.
(b) Illustration of a SyAF $2$-layer skyrmion tube (i.e., $N=2$). Black arrows indicate the Magnus force acting on each FM layer.
(c) Illustration of a SyAF $6$-layer skyrmion tube (i.e., $N=6$).
(d) Definitions of $R_x$ and $R_y$, which are used to describe the size and shape of the skyrmion in the $x-y$ plane of each FM layer.
}
\label{FIG1}
\end{figure}

\section{Methods}
\label{se:Modeling}

Figure~\ref{FIG1}(a) illustrates the SyAF multilayer nanotracks. The SyAF $N$-layer nanotrack ($N\geq2$) includes $N$ FM layers, which are strictly exchange coupled in an antiferromagnetic (AFM) manner by interlayer AFM exchange interactions. In all SyAF multilayer nanotracks, the length along the $x$ direction, the width along the $y$ direction, and the thickness along the $z$ direction are equal to $100$ nm, $100$ nm, and $12$ nm, respectively. The periodic boundary conditions are applied in the $x$ and $y$ directions. It should be mentioned that two adjacent FM layers should be separated by a nonmagnetic metal spacer in real experimental samples; however, we ignore the thickness of the nonmagnetic spacer but preserve the effect of the nonmagnetic spacer in the simulation for the sake of simplicity, which saves computational power.

In this work, we explicitly consider the SyAF $N$-layer nanotracks with $N=2,4,6,12$. For the SyAF multilayer nanotrack with $N=2$, two $6$-nm FM layers are antiferromagnetically exchange coupled. For the SyAF multilayer nanotrack with $N=4$, four $3$-nm FM layers are antiferromagnetically exchange coupled. For the SyAF multilayer nanotrack with $N=6$, six $2$-nm FM layers are antiferromagnetically exchange coupled. For the SyAF multilayer nanotrack with $N=12$, 12 $1$-nm FM layers are antiferromagnetically exchange coupled.
At the initial state, the skyrmion tube is relaxed at the position of $x=50$ nm, $y=50$ nm. The total skyrmion number $Q_{\text{tot}}$ of the SyAF $N$-layer skyrmion tube is equal to zero due to the nature of the SyAF nanotrack~\cite{Xichao_PRB2016B}. We consider a current-perpendicular-to-plane geometry, where the driving spin current is injected into all FM layers vertically.

The total Hamiltonian $H$ is decomposed into the Hamiltonian for each FM layer $H_{n}$ and the interlayer AFM exchange coupling $H_{\text{inter}}$ between neighboring FM layers,
\begin{equation}
H=\sum_{n=1}^{N}H_{n}+H_{\text{inter}}.
\label{eq:Hamil-total}
\end{equation}
The Hamiltonian for each FM layer reads
\begin{eqnarray}
H_{n}&=&
-A_{\text{intra}}\sum_{\langle i,j\rangle}\boldsymbol{m}_{i}^{n}\cdot\boldsymbol{m}_{j}^{n}+K\sum_{i}\left[1-\left(m_{i}^{n,z}\right)^{2}\right] \nonumber \\
&&+D_{ij}\sum_{\langle i,j\rangle}\left(\boldsymbol{\nu}_{ij}\times\hat{z}\right)\cdot\left(\boldsymbol{m}_{i}^{n}\times\boldsymbol{m}_{j}^{n}\right)+H_{\text{DDI}},
\label{eq:Hamil-intralayer}
\end{eqnarray}
where $n$ is the FM layer index ($n=1,2,\cdots,N$), $\boldsymbol{m}_{i}^{n}$ represents the local magnetic moment orientation normalized as $|\boldsymbol{m}_{i}^{n}|=1$ at site $i$, and $\left\langle i,j\right\rangle$ runs over all the nearest-neighbor sites in each FM layer. The first term represents the intralayer FM exchange interaction with the intralayer FM exchange stiffness $A_{\text{intra}}$. The second term represents the DM interaction (DMI), where $D_{ij}$ is the DMI coupling energy and $\boldsymbol{\nu}_{ij}$ is the unit vector between sites $i$ and $j$. The third term represents the perpendicular magnetic anisotropy with the anisotropy constant $K$. $H_{\text{DDI}}$ represents the dipole-dipole interaction. The Hamiltonian for the interlayer AFM exchange interactions reads
\begin{equation}
H_{\text{inter}}=-\sum_{n=1}^{N-1}A_{\text{inter}}\sum_{i}\boldsymbol{m}_{i}^{n}\cdot\boldsymbol{m}_{i}^{n+1}.
\label{eq:Hamil-interlayer}
\end{equation}
Here the interlayer exchange stiffness $A_{\text{inter}}$ is negative due to the interlayer AFM exchange interaction.

For the current-induced dynamics, we numerically solve the Landau-Lifshitz-Gilbert equation including the damping-like and fieldlike spin-orbit torques (SOTs), given as~\cite{OOMMF,Xichao_PRB2016B,Tomasello_JPD2017}
\begin{align}
\frac{d\boldsymbol{m}}{dt}=&-\gamma_{0}\boldsymbol{m}\times\boldsymbol{h}_{\text{eff}}+\alpha\left(\boldsymbol{m}\times\frac{d\boldsymbol{m}}{dt}\right)\notag \\
&-u\boldsymbol{m}\times\left(\boldsymbol{m}\times\boldsymbol{p}\right)-\xi u\left(\boldsymbol{m}\times\boldsymbol{p}\right).
\label{eq:LLGS}
\end{align}
Here, $\boldsymbol{h}_{\text{eff}}=-\frac{1}{\mu_0 M_\text{S}}\cdot\frac{\partial H}{\partial\mathbf{m}}$ is the effective field.
$\mu_0$ is the vacuum permeability constant, and $M_\text{S}$ is the saturation magnetization.
$\gamma_{0}$ is the gyromagnetic ratio with its absolute value, and $\alpha$ is the Gilbert damping coefficient.
$u=|\frac{\gamma_{0}\hbar}{\mu_{0}e}|\frac{j\theta_{\text{SH}}}{2aM_{\text{S}}}$ is the dampinglike SOT coefficient, and $\xi$ is the relative strength of the fieldlike torque.
$\boldsymbol{p}=-y$ represents the unit spin polarization vector, $\hbar$ is the reduced Planck constant, $e$ is the electron charge, $j$ is the applied driving current density, $\theta_{\text{SH}}=0.1$ is the spin Hall angle, and $a=1$ nm is the thickness of the cell size.

The simulation is performed by using the 1.2a5 release of the Object Oriented MicroMagnetic Framework (OOMMF) developed at NIST~\cite{OOMMF}.
The simulation uses the OOMMF extensible solver (OXS) objects of the standard OOMMF distribution along with the OXS extension modules for the interface-induced DMI~\cite{OOMMFDMI,Rohart_PRB2013}.
The cell size used in the simulation is $2$ nm $\times$ $2$ nm $\times$ $1$ nm, which guarantees both numerical accuracy and computational efficiency.
The magnetic parameters used in the simulation are~\cite{Fert_NNANO2013,Sampaio_NNANO2013,Tomasello_SREP2014,Xichao_NCOMMS2016,Xichao_PRB2016B}: $\alpha=0.01-0.5$ with a default value of $0.1$, ${\gamma_0}=2.211\times 10^{5}$ m/(As), $M_{\text{S}}=1000$ kA/m, $A_{\text{intra}}=10$ pJ/m, $A_{\text{inter}}=-1$ pJ/m (i.e., $\sigma=-1$ mJ/m$^2$), $D=1.1$ mJ/m$^2$ (for $N=2$), $D=1.3$ mJ/m$^2$ (for $N>2$), and $K=0.8$ MJ/m$^{3}$.

\section{Results and Discussions}
\label{se:motion}

\begin{figure}[!t]
\centerline{\includegraphics[width=0.490\textwidth]{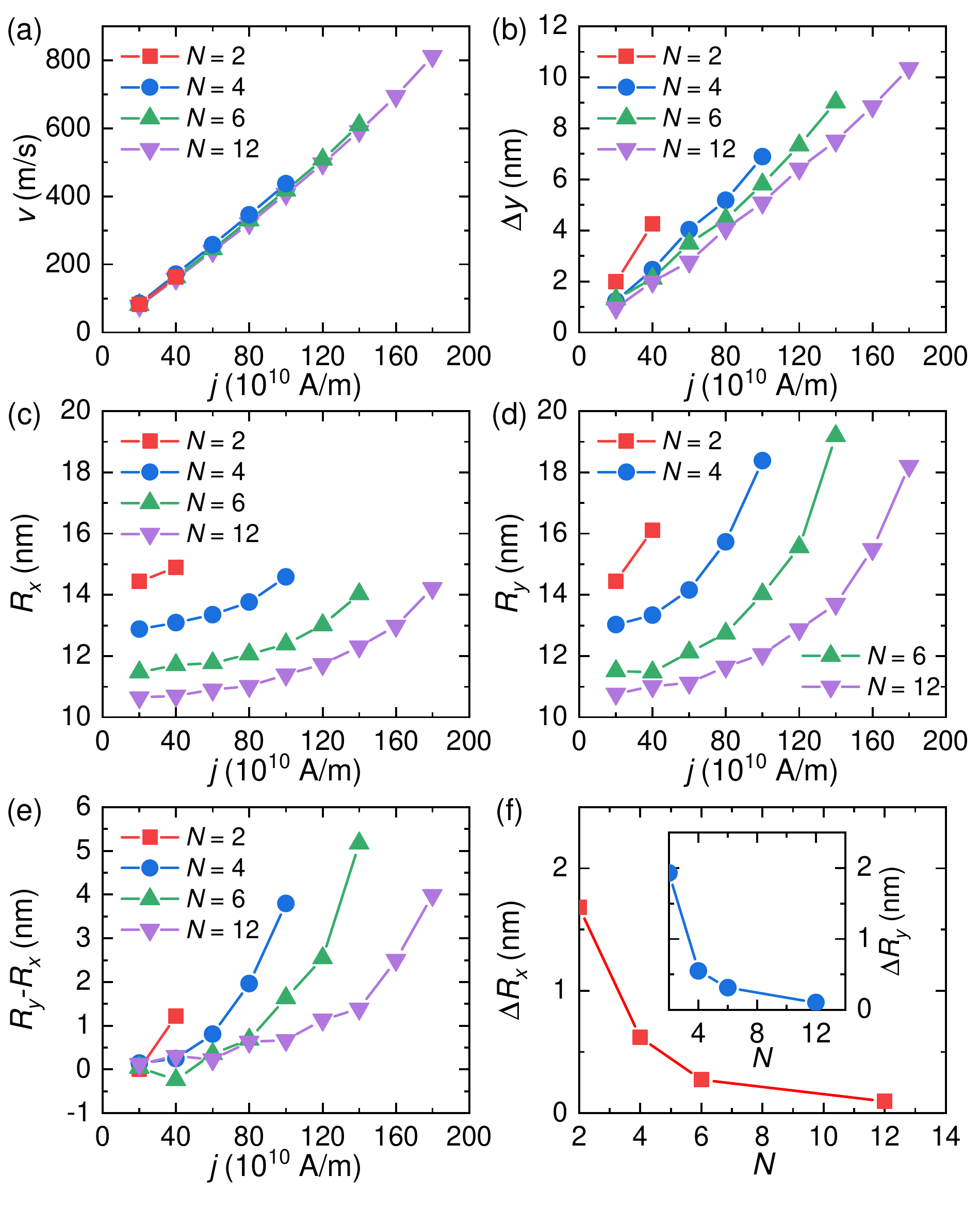}}
\caption{%
(a) Skyrmion tube velocity $v$ as a function of driving current density $j$ for a SyAF $N$-layer skyrmion.
(b) Horizontal distance between the top layer and bottom layer skyrmion centers in the $y$ direction $\Delta y$ as a function of driving current density $j$. Note that when the SyAF $N$-layer skyrmion is driven into motion in the $x$ direction, the velocities of skyrmions in each layer are the same. Thus, the skyrmion center positions in the $x$ direction are the same in all FM layers, i.e., $\Delta x=0$.
(c) $R_x$ as a function of driving current density $j$ for the skyrmion in the bottom FM layer.
(d) $R_y$ as a function of driving current density $j$ for the skyrmion in the bottom FM layer.
(e) $R_y-R_x$ as a function of driving current density $j$ for the skyrmion in the bottom FM layer.
(f) $\Delta R_x$ (i.e., $R_x^\mathrm{bottom}-R_x^\mathrm{top}$) as a function of $N$ when $j=20\times 10^{10}$ A/m.
The inset shows the corresponding $\Delta R_y$ (i.e., $R_y^\mathrm{bottom}-R_y^\mathrm{top}$).
}
\label{FIG2}
\end{figure}

We start with a computational investigation of the current-velocity relation of the skyrmion tubes in SyAF $N$-layer nanotracks with $N=2,4,6,12$, where we initially consider only the dampinglike torque (i.e., $\xi=0$).
It is found that the velocity of the skyrmion tube is proportional to the driving current density, as shown in Fig.~\ref{FIG2}(a).

For the steady motion of the rigid skyrmion tubes in SyAF $N$-layer nanotracks, we use the Thiele equation~\cite{Thiele_PRL1973,Tomasello_SREP2014} to interpret the simulation results. The Thiele equation for the skyrmion in each FM layer reads
\begin{equation}
\boldsymbol{G}^{n}\times\boldsymbol{v}^{n}-\alpha\boldsymbol{\mathcal{D}}^{n}\cdot\boldsymbol{v}^{n}+\boldsymbol{p}\cdot\boldsymbol{\mathcal{B}}^{n}+\boldsymbol{F}^{n}=\boldsymbol{0},
\label{eq:ThieleEq_layer}
\end{equation}
with $n$ being the layer index. $\boldsymbol{\mathcal{D}}^{n}$, $\mathbf{v}^{n}$, $\boldsymbol{\mathcal{B}}^{n}$, and $\boldsymbol{F}^{n}$ represent the dissipative tensor, the skyrmion velocity, the tensor related to the driving current, and the effective force due to the AFM interlayer exchange coupling, respectively.
$\mathbf{G}^{n}=T^{n}\frac{M_\text{S}}{\gamma_0}(0,0,Q^{n})$ is the gyromagnetic coupling constant representing the Magnus force, with $Q^{n}$ being the skyrmion number, where $T^{n}$ is the thickness of the FM sublayer.
It should be noted that the Thiele equation [i.e., Eq.~(\ref{eq:ThieleEq_layer})] essentially does not include the thickness for the two-dimensional model as the contributions of the thickness are the same in all terms.
The skyrmion number in each FM layer is defined as
\begin{equation}
Q^{n}=-\frac{1}{4\pi}\int\boldsymbol{m}^{n}\cdot\left(\partial_{x}\boldsymbol{m}^{n}\times\partial_{y}\boldsymbol{m}^{n}\right)dxdy.
\label{eq:SkNum}
\end{equation}
We have taken the same damping coefficient $\alpha$ for all FM layers.
$\mathbf{\mathcal{D}}^{n}$ is the dissipative tensor, with
$\mathcal{D}^{n}_{\mu\nu}=T^{n}\frac{M_\text{S}}{\gamma_0}\int\partial_{\mu}\boldsymbol{m}^{n}\cdot\partial_{\nu}\boldsymbol{m}^{n}\,dxdy/{4\pi}$.
$\boldsymbol{\mathcal{B}}^{n}$ is the tensor related to the driving force, with $\mathcal{B}^{n}_{\mu\nu}=-T^{n}\frac{M_\text{S}}{\gamma_0}u\int\left(\partial_{\mu}\boldsymbol{m}^{n}\times\boldsymbol{m}^{n}\right)_{\nu}\,dxdy/{4\pi}$.

First, we assume that all sublayer skyrmions of a skyrmion tube move together with the same velocity $\mathbf{v}$ since they are tightly bound in an AFM configuration.
Summing all $n$ Thiele equations~(\ref{eq:ThieleEq_layer}), we can phenomenologically obtain
\begin{equation}
-\alpha\boldsymbol{\mathcal{D}}\cdot\boldsymbol{v}+\boldsymbol{p}\cdot\boldsymbol{\mathcal{B}}=\mathbf{0},
\label{eq:ThieleEq-total}
\end{equation}
where the interlayer AFM forces are canceled out, i.e., $\sum\mathbf{F}^{n}=0$. The Magnus forces are also canceled out, i.e., $\sum\mathbf{G}^{n}=0$.
Solving Eq.~(\ref{eq:ThieleEq-total}), the velocity of the SyAF skyrmion tube can be obtained
\begin{equation}
v_x=\frac{uI}{\alpha\mathcal{D}},\,\,\,\,\, v_y=0,
\label{eq:velocity}
\end{equation}
where $I=\pi r_\text{sk}/4$ and $\mathcal{D}=\pi^2/8$.
The theoretical solutions show that the skyrmions in each FM layer steadily move along the $x$ direction given that they are strictly exchange-coupled antiferromagnetically. The skyrmion velocity is proportional to the driving force, which is in line with the simulation results.

As shown in Fig.~\ref{FIG2}(a), the dynamic stability of the SyAF skyrmion tube is enhanced when the number of FM layers increases.
For example, the SyAF $2$-layer skyrmion tube is destroyed when the driving current density $j>40\times 10^{10}$ A/m. The SyAF $4$-layer skyrmion tube is destroyed when $j>100\times 10^{10}$ A/m. The SyAF $6$-layer skyrmion tube is destroyed when $j>140\times 10^{10}$ A/m. The SyAF $12$-layer skyrmion tube is destroyed when $j>180\times 10^{10}$ A/m.
The critical current density above which the skyrmion tube is destroyed increases when the number of layers increases. It should be noted that the pinning in materials could help stabilize the skyrmion tube for the large driving current density~\cite{Reichhardt_RPP2016,Reichhardt_arxiv2021}. In addition, the critical current density decreases as the strength of the interlayer AFM exchange coupling decreases. When the strength of the interlayer AFM exchange coupling decreases, the skyrmions can be more easily decoupled and destroyed due to the interaction between the skyrmion and the sample edge.

The destruction of the moving skyrmion tube is caused by the fact that the Magnus forces acting on sublayer skyrmions with opposite skyrmion numbers $Q^{n}$ point in opposite directions, which may deform and pull apart the skyrmion tube when the Magnus forces are larger than a certain threshold.
The magnitude of the Magnus force [i.e., $\boldsymbol{G}^{n}\times\boldsymbol{v}^{n}$; see Eq.~(\ref{eq:ThieleEq_layer})] is proportional to the skyrmion speed as well as the magnetization and sublayer thickness~\cite{Seki_BOOK2016}, which can be seen from the definition
\begin{align}
G^{n}=&T^{n}\frac{M_\text{S}}{\gamma_0}Q^{n} \notag \\
=&-T^{n}\frac{M_\text{S}}{\gamma_0} \frac{1}{4\pi}\int\boldsymbol{m}^{n}\cdot\left(\partial_{x}\boldsymbol{m}^{n}\times\partial_{y}\boldsymbol{m}^{n}\right)dxdy,
\label{eq:GV}
\end{align}
where $T^{n}$ is the thickness of the FM sublayer. Hence, it can be seen that in the SyAF multilayers with identical total thicknesses, the skyrmion tube with fewer layers (i.e. smaller $N$) could more easily be deformed by the Magnus force.
To be more specific, the Magnus force will lead to the shift of sublayer skyrmions in the $\pm y$ directions.
Due to the Magnus-force-induced deformation, the SyAF skyrmion tube velocities are slightly different for the SyAF nanotracks with different $N$, especially when the driving current density is large.

\begin{figure}[t]
\centerline{\includegraphics[width=0.490\textwidth]{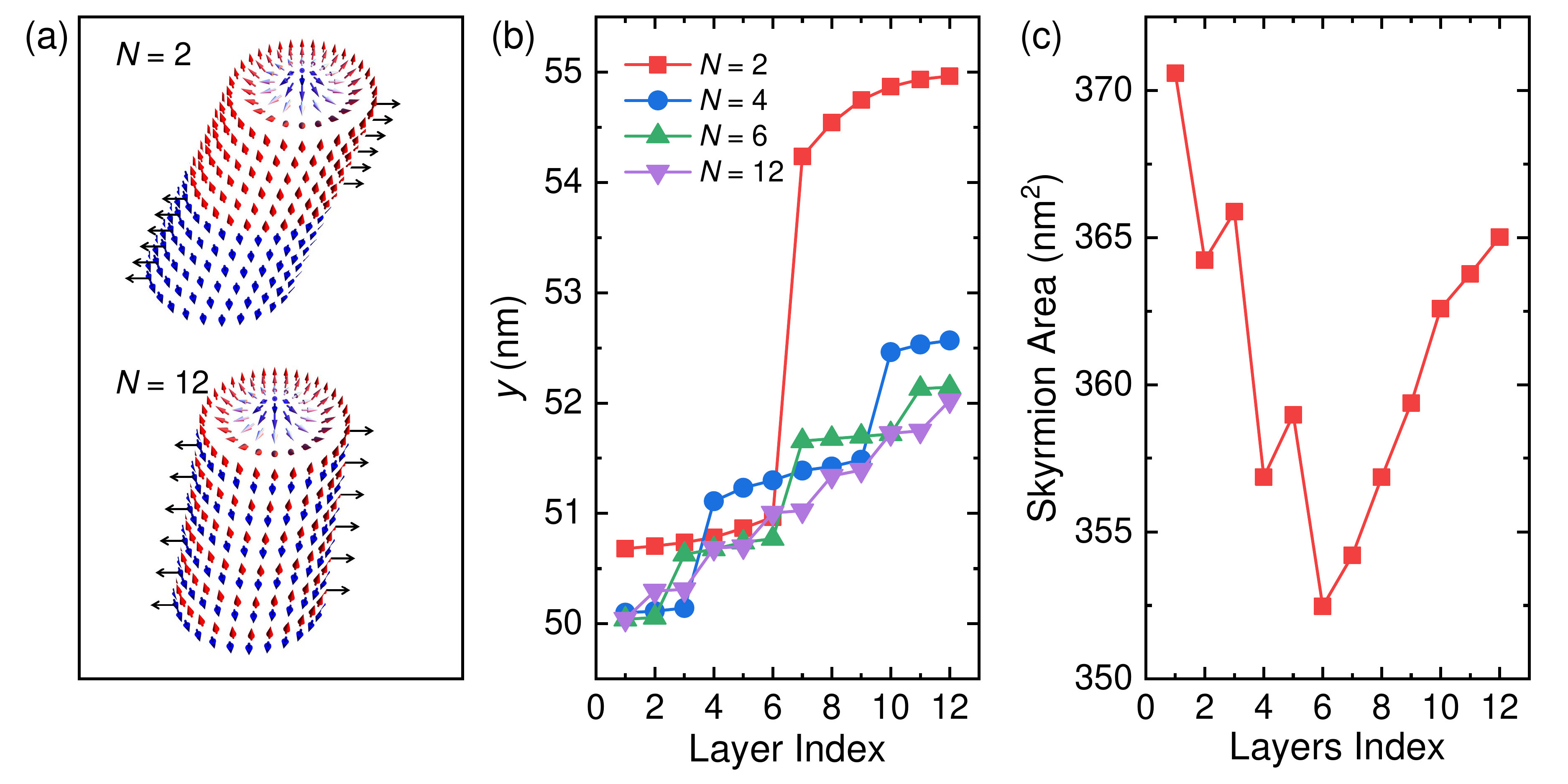}}
\caption{%
(a) Schematic illustrations of deformed moving SyAF skyrmion tubes. The total thickness is $12$ nm, i.e., $12$ spins in the thickness direction. $N$ denotes the number of FM layers in a sample. For $N=2$, the thickness of each FM layer equals $6$ nm. For $N=12$, the thickness of each FM layer equals $1$ nm. At the same driving current density, the Magnus-force-induced deformation of the SyAF $2$-layer skyrmion tube is larger than that of the SyAF $12$-layer skyrmion.
(b) Sublayer skyrmion center locations (in the $y$ direction) of deformed SyAF $N$-layer skyrmion tubes driven by a current density of $j=40\times 10^{10}$ A/m. The layer index indicates the single-spin-thick sublayer position, for example, $1$ and $12$ denote the bottommost and topmost layers of the SyAF structure.
(c) Sublayer skyrmion areas of a deformed SyAF $12$-layer skyrmion tube driven by a current density of $j=40\times 10^{10}$ A/m.
}
\label{FIG3}
\end{figure}

Figure~\ref{FIG2}(b) shows the distance (i.e., $\Delta y$) in the $y$ direction between the top sublayer and bottom sublayer skyrmions as a function of the driving current density.
$\Delta y$ increases with increasing driving current density. When the driving current density increases, the Magnus force acting on skyrmions in each FM layer increases, leading to a larger shift of sublayer skyrmion centers.
However, $\Delta y$ decreases when the number of FM layers (i.e., $N$) increases at a given driving current density.
For example, when $j=100\times 10^{10}$ A/m, $\Delta y=7$ nm for the SyAF $4$-layer skyrmion, and $\Delta y$ decreases to $5$ nm for the SyAF $12$-layer skyrmion. Note that the total thickness of the SyAF nanotracks is fixed at $12$ nm.

We further investigate the deformation of SyAF skyrmion tubes.
The geometries of bottom sublayer skyrmions are described by $R_x$, $R_y$, and $R_y-R_x$ in Figs.~\ref{FIG2}(c)-\ref{FIG2}(e).
The sublayer skyrmions of a moving SyAF skyrmion tube are elongated in the $y$ direction.
The deformation is significant when the driving current density is large because the Magnus force [i.e., $\boldsymbol{G}^{n}\times\boldsymbol{v}^{n}$; see Eq.~(\ref{eq:ThieleEq_layer})] acting on each FM sublayer increases with the current-induced velocity.
However, it can be seen that the deformation of the SyAF $12$-layer skyrmion tube is smaller than that of the SyAF $4$-layer and $6$-layer skyrmion tubes when $j>80\times 10^{10}$ A/m.
The reason is that the Magnus force also decreases with decreasing thickness of the FM sublayer [see Eq.~(\ref{eq:GV})]. For the SyAF $4$-layer skyrmion tube, the thickness of each FM sublayer equals $3$ nm, while it is equal to $1$ nm for the SyAF $12$-layer skyrmion tube.

We also study the geometries of sublayer skyrmions in the topmost and bottommost FM layers. Figure~\ref{FIG2}(f) shows $\Delta R_x$ (i.e., $R_x^\mathrm{bottom}-R_x^\mathrm{top}$) and $\Delta R_y$ (i.e., $R_y^\mathrm{bottom}-R_y^\mathrm{top}$) as functions of $N$. For the SyAF $2$-layer skyrmion tube, $\Delta R_x$ and $\Delta R_y$ are about $2$ nm. For the SyAF $12$-layer skyrmion tube, $\Delta R_x$ and $\Delta R_y$ are almost zero.
The reason behind this phenomenon could be the effect of the dipole-dipole interaction. Namely, when the thickness of FM layers is large, the dipole-dipole interaction may result in a certain nonuniformity and tilt of the skyrmion tube in the thickness direction.
Note that we do not observe the helicity oscillation of the skyrmions, which may be caused by complex stray field interactions at certain conditions~\cite{Lemesh_PRApplied2019}. In our SyAF structures, $M_\text{S}$ of all FM layers are the same; therefore, there is no stray field in the system.

In Fig.~\ref{FIG3}(a), we illustrate two deformed SyAF skyrmion tubes driven by a current density of $j=40\times 10^{10}$ A/m.
The slanted deformation of the SyAF $2$-layer skyrmion tube is obviously larger than that of the SyAF $12$-layer skyrmion tube.
For the SyAF $2$-layer skyrmion tube, the Magnus forces acting on the top FM and bottom FM layers are large (due to the thickness of the FM sublayers) and point in opposite directions, which leads to the deformation of the skyrmion tube along the direction of Magnus forces (i.e., the $\pm y$ directions).
In contrast, for the SyAF $12$-layer skyrmion tube, the magnitude of Magnus forces is much smaller due to the reduced thickness of each FM sublayer. At the same time, the Magnus forces acting on $12$ FM sublayers are opposite to each other in a staggered manner, which leads to a better cancellation of Magnus forces and smaller deformation of the SyAF skyrmion tube.
As shown in Fig.~\ref{FIG3}(b), for the SyAF multilayer with a given total thickness of $12$ nm, the current-induced deformation of the SyAF $N$-layer skyrmion tube in the Magnus force direction (i.e., the $\pm y$ directions) driven by $j=40\times 10^{10}$ A/m decreases with increasing number of FM sublayers. Namely, the deformation decreases with decreasing thickness of the FM sublayers.
For the case of $N=2$, the horizontal spacing between the topmost and bottommost sublayer skyrmions equals $\sim 4$ nm, while it equals $\sim 2$ nm for the case of $N=12$.

On the other hand, it is worth mentioning that the large leap of the $N=2$ case in Fig.~\ref{FIG3}(b) indicates that the slanted deformation of the SyAF $2$-layer skyrmion tube is most significant at the antiferromagnetically exchange coupled interface, where the shear strain is maximum from a phenomenological point of view. However, for other cases with $N>2$, the reduced Magnus forces as well as increased number of antiferromagnetically exchange coupled interfaces cannot lead to obvious shear strain (i.e., leaps) at interfaces.

Note that, as mentioned above [see Fig.~\ref{FIG2}(f)], the sublayer skyrmion size is not uniform in the thickness direction, as shown in Fig.~\ref{FIG3}(c), which may be caused by complex dipole-dipole interactions in the SyAF multilayer structure.
For example, the size of the sublayer skyrmion is larger near the top and bottom multilayer surfaces for the SyAF $12$-layer skyrmion tube, while it is smaller in the mid interior of the multilayer.
In particular, the sublayer skyrmion size in the bottommost layer is larger than that in the topmost layer.
As the magnitude of Magnus force acting on each sublayer FM skyrmion is also proportional to the sublayer skyrmion size (i.e., in addition to the sublayer thickness), the nonuniformity and asymmetry of the SyAF skyrmion tube in the thickness direction may result in the fact that the Magnus forces cannot be canceled perfectly, especially during the acceleration of the SyAF skyrmion tube upon the application of driving current.
Consequently, the uncompensated Magnus forces may lead to complex dynamic deformation and a transverse shift of the SyAF skyrmion tube.
Namely, when the SyAF skyrmion tube reaches steady motion, it may show certain deformation in three dimensions as well as a certain transverse shift of its average center in the $\pm y$ directions, which are most significant for the case of $N=2$ [see Fig.~\ref{FIG3}(b)].

\begin{figure}[t]
\centerline{\includegraphics[width=0.490\textwidth]{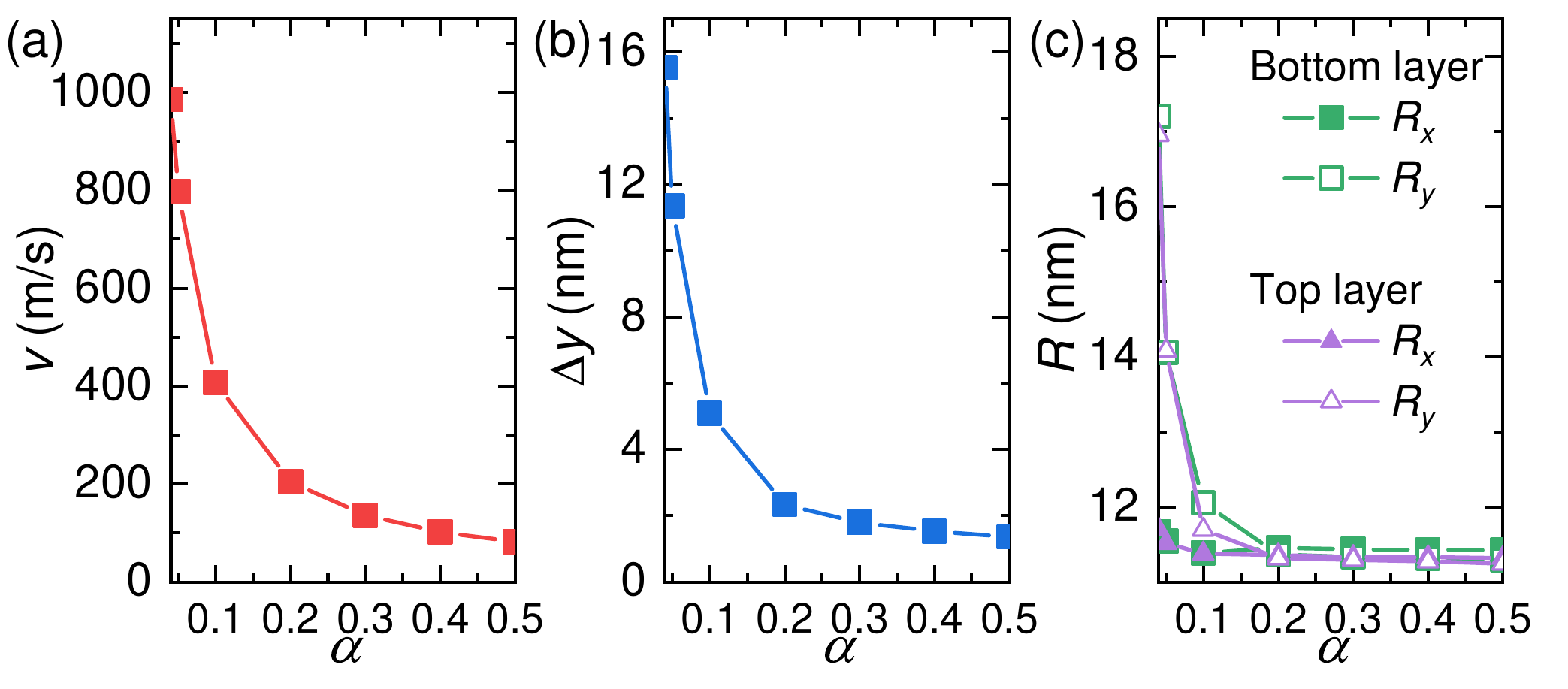}}
\caption{%
(a) Damping dependence of the SyAF $12$-layer skyrmion tube velocity $v$ at $j=100\times 10^{10}$ A/m.
(b) Damping dependence of $\Delta y$ of the SyAF $12$-layer skyrmion tube at $j=100\times 10^{10}$ A/m.
(c) Damping dependences of $R_x$ and $R_y$ of the SyAF $12$-layer skyrmion tube at $j=100\times 10^{10}$ A/m.
}
\label{FIG4}
\end{figure}

\begin{figure}[t]
\centerline{\includegraphics[width=0.490\textwidth]{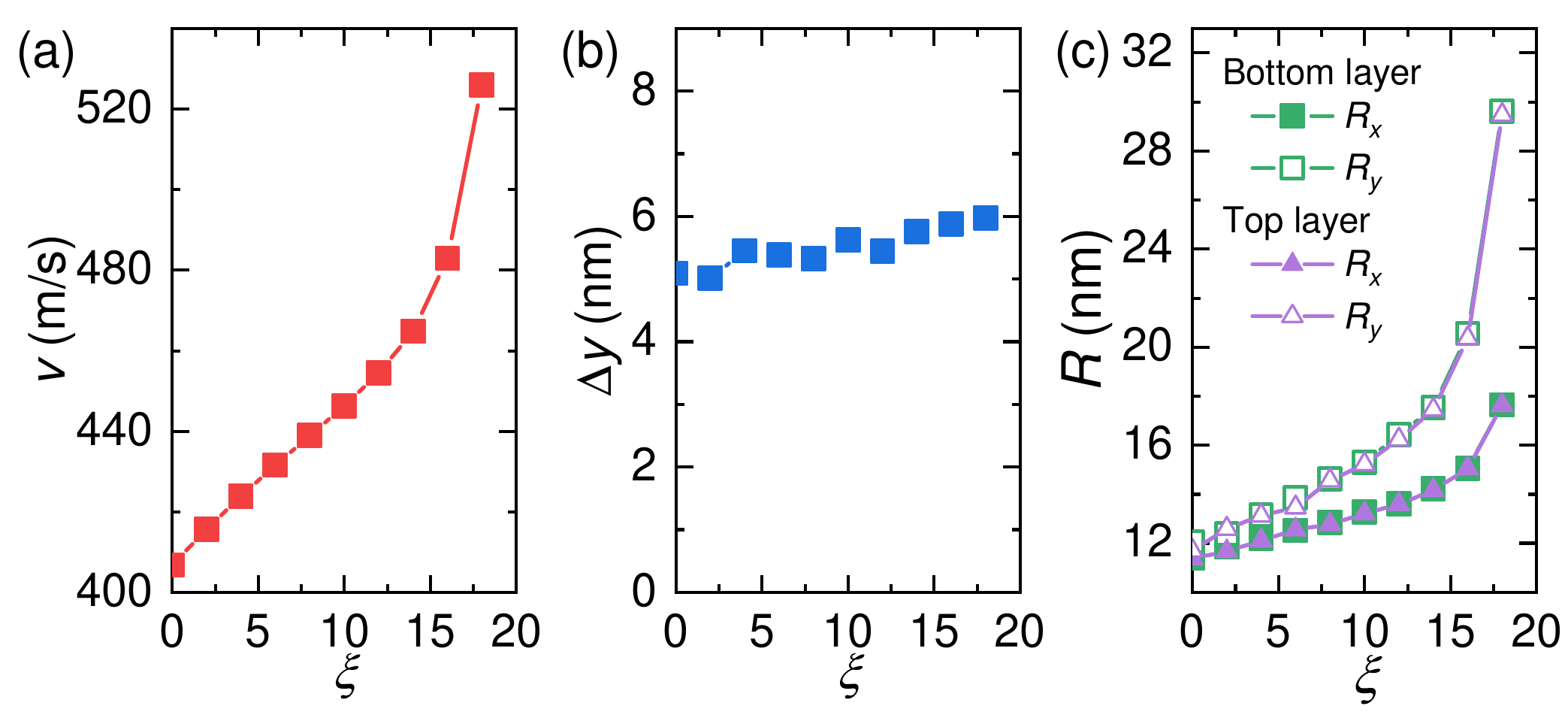}}
\caption{%
Effect of the field-like torque strength $\xi$ on the current-induced motion of a SyAF $12$-layer skyrmion at $j = 100 \times 10^{10}$~A/m.
(a) Velocity, (b) $\Delta y$, and (c) $R_x$ and $R_y$ as a function of $\xi$.
}
\label{FIG5}
\end{figure}

The effect of the damping parameter $\alpha$ on the current-induced motion of the SyAF skyrmion tube is also investigated.
Figure~\ref{FIG4} shows the results for the current-induced motion of a SyAF $12$-layer skyrmion tube, which is the most stable SyAF skyrmion tube studied in this work.
The skyrmion tube velocity decreases with increasing $\alpha$ [see Fig.~\ref{FIG4}(a)], which follows the theoretical solution given in Eq.~(\ref{eq:velocity}).
The shift of the sublayer skyrmion centers in the $y$ direction also decreases when $\alpha$ increases, as shown in Fig.~\ref{FIG4}(b).
Figure~\ref{FIG4}(c) shows $R_x$ and $R_y$ of sublayer skyrmions in the topmost and bottommost FM layers. When $\alpha=0.04$, the deformation of sublayer skyrmions both in top and bottom FM layers are significant, where $R_y-R_x$ reaches $5$ nm. When $\alpha=0.5$, $R_x$ and $R_y$ are almost identical, indicating an insignificant distortion.
In this work, we consider only the case where the damping parameter $\alpha$ is the same in all FM layers. For the case where $\alpha$ are different in different FM layers, the skyrmions may still be coupled tightly when the driving current density is small. However, when the driving current density is large, the skyrmions may be decoupled due to the $\alpha$-induced differences in Magnus force and motion direction of different skyrmions. Note that the critical driving current density above which the skyrmions are decoupled increases when $\alpha$ increases~\cite{Reichhardt_PRE2019}. In addition, it is worth mentioning that the inhomogeneous driving current in SyAF multilayers could also lead to a decoupling transition, which is similar to the transformer effect in layered superconductors~\cite{Giaever_PRL1965}.

We also study the effect of the fieldlike torque on the current-induced motion of a SyAF $12$-layer skyrmion.
Figure~\ref{FIG5}(a) shows the velocity of the skyrmion tube as a function of the fieldlike torque strength $\xi$.
The fieldlike torque can increase the size of sublayer skyrmions, which results in the rise of the skyrmion tube velocity as the skyrmion velocity is proportional to the skyrmion size at a given current density~\cite{Xichao_PRB2016C}.
The shift of the sublayer skyrmion centers in the $y$ direction slightly increases with increasing $\xi$, as shown in Fig.~\ref{FIG5}(b).
The fieldlike torque can also lead to the expansion of sublayer skyrmions as well as the deformation of skyrmion tube [see Fig.~\ref{FIG5}(c)].

\begin{figure}[t]
\centerline{\includegraphics[width=0.490\textwidth]{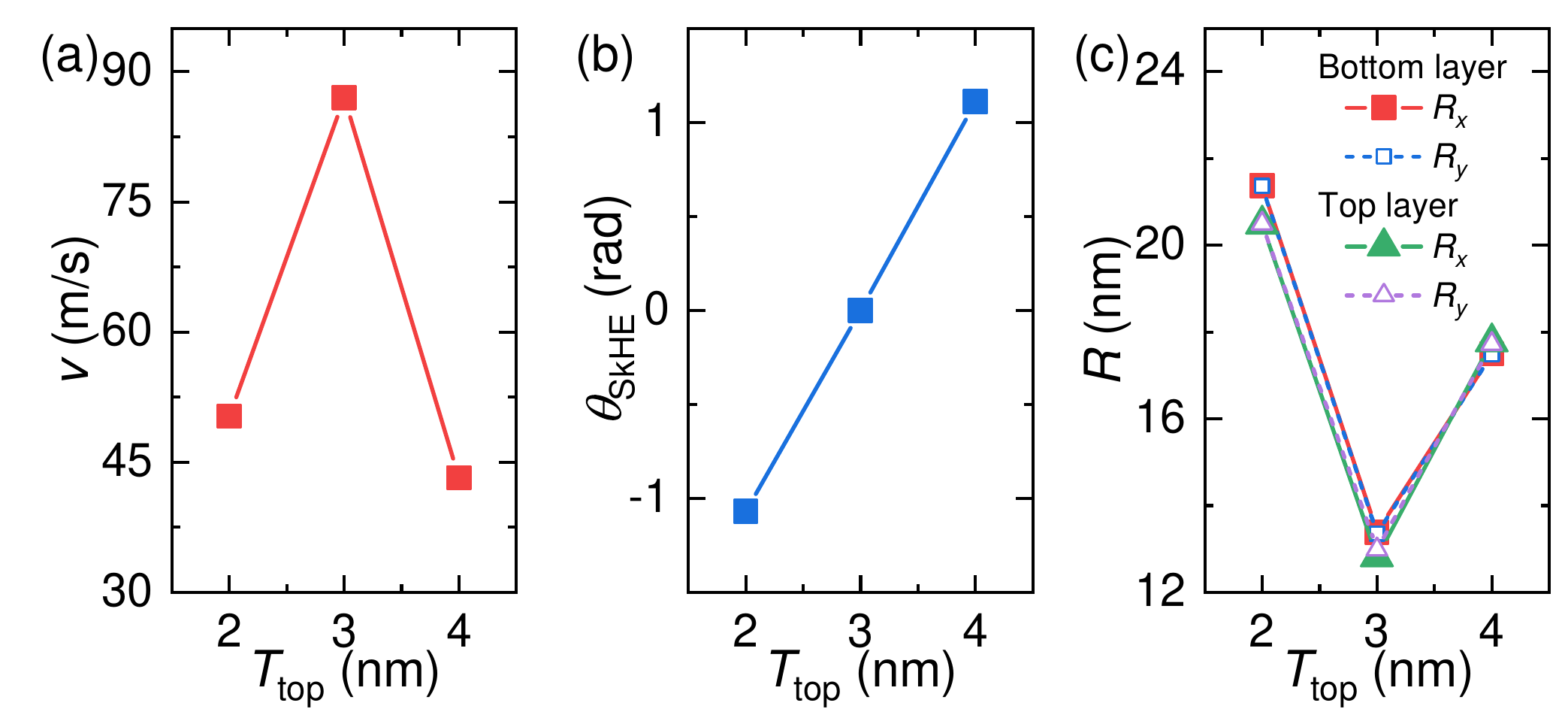}}
\caption{%
The current-induced motion of a SyAF bilayer skyrmion (i.e., $N=2$). Here the total thickness of the sample is fixed at $6$ nm. The thicknesses of the top and bottom FM layers are defined as $T_\mathrm{top}$ and $T_\mathrm{bottom}$, respectively. Namely, $T_\mathrm{top}+T_\mathrm{bottom}=6$ nm.
(a) Velocity, (b) skyrmion Hall angle $\theta_\mathrm{SkHE}$, and (c) $R_x$ and $R_y$ as  function of $T_\mathrm{top}$ at $j=20\times 10^{10}$ A/m.
}
\label{FIG6}
\end{figure}

In the above simulations we assume a fixed thickness of each FM layer. Here we proceed to investigate the effect of sublayer thickness $T$ on the skyrmion tube dynamics, as shown in Fig.~\ref{FIG6}.
In this part, we consider a SyAF bilayer nanotrack (i.e., $N=2$) with a fixed total thickness of $6$ nm (i.e., $T_\text{top}+T_\text{bottom}=6$ nm).
We simulate three cases, i.e., $T_\text{top}=2$, $3$, and $4$ nm.
Figure~\ref{FIG6}(a) shows the current-driven motion of the SyAF bilayer skyrmion tube. Due to the AFM exchange coupling, the sublayer skyrmions in top and bottom FM layers are exchange coupled tightly and move together.
When $T_\mathrm{top}=T_\mathrm{bottom}=3~\mathrm{nm}$, the velocity reaches $87$ m/s, and the skyrmion Hall angle is equal to zero [see Fig.~\ref{FIG6}(b)].
When $T_\mathrm{top}\neq T_\mathrm{bottom}$, the skyrmion tube velocity is reduced, and the skyrmion tube shows the skyrmion Hall effect.
As shown in Fig.~\ref{FIG6}(c), the skyrmion tube deformation increases when $T_\mathrm{top}\neq T_\mathrm{bottom}$.

\section{Conclusion}
\label{se:Conclusions}

In conclusion, we have studied the current-induced motion of skyrmion tubes in SyAF multilayer nanotracks.
The SyAF $N$-layer skyrmion tubes consist of $N$ sublayer FM skyrmions, which are strictly exchange coupled antiferromagnetically.
It is found that for SyAF $N$-layer multilayers with identical total thicknesses, the current-driven dynamic stability of the SyAF skyrmion tube increases with increasing $N$. As a result, the SyAF $N$-layer skyrmion with a higher $N$ can be driven by a larger current density and thus, can reach a higher speed.
Furthermore, we have studied the effects of the damping parameter and fieldlike torque on the moving SyAF $N$-layer skyrmion tube.
When the damping parameter is large, the motion of a SyAF $N$-layer skyrmion will be more stable, while its speed will be reduced.
The fieldlike torque can deform the SyAF skyrmion tube, but it can also lead to a speed increase of the SyAF skyrmion tube.
In addition, we computationally demonstrated the effect of sublayer thickness on the skyrmion Hall effect of a SyAF bilayer skyrmion tube.
For the SyAF bilayer skyrmion, when the thicknesses of the top and bottom FM layers are identical, the SyAF skyrmion shows no skyrmion Hall effect due to the cancellation of the Magnus forces.
However, when the thicknesses of the top and bottom FM layers are different, the skyrmion Hall effect cannot be eliminated.
We believe our results are useful for understanding the dynamic stability and mobility of the skyrmion tubes in SyAF structures. We also believe our results can provide guidelines for building SyAF spintronic devices based on topological spin textures.

\begin{acknowledgments}
X.Z. was an International Research Fellow of Japan Society for the Promotion of Science (JSPS).
X.Z. was supported by JSPS KAKENHI (Grant No. JP20F20363).
M.E. acknowledges the support by the Grants-in-Aid for Scientific Research from JSPS KAKENHI (Grant Nos. JP18H03676 and JP17K05490) and the support by CREST, JST (Grant Nos. JPMJCR20T2 and JPMJCR16F1).
O.A.T. acknowledges the support by the Australian Research Council (Grant No. DP200101027), the Cooperative Research Project Program at the Research Institute of Electrical Communication, Tohoku University (Japan), and by the NCMAS grant.
X.L. acknowledges the support by the Grants-in-Aid for Scientific Research from JSPS KAKENHI (Grant Nos. JP20F20363 and JP21H01364).
G.Z. acknowledges the support by the National Natural Science Foundation of China (Grant Nos. 52111530143 and 51771127), and the Scientific Research Fund of Sichuan Provincial Education Department (Grant Nos. 18TD0010 and 16CZ0006).
Y.Z. acknowledges the support by the Guangdong Special Support Project (Grant No. 2019BT02X030), Shenzhen Peacock Group Plan (Grant No. KQTD20180413181702403), Pearl River Recruitment Program of Talents (Grant No. 2017GC010293), and National Natural Science Foundation of China (Grant Nos. 11974298 and 61961136006).
\end{acknowledgments}


\end{document}